\documentclass[prl,twocolumn,preprintnumbers,amsmath,amssymb,floatfix,superscriptaddress,showpacs]{revtex4}
\usepackage{graphicx}
\usepackage{graphics}
\usepackage{psfrag}
\usepackage{hyperref}
\usepackage{multirow}
\usepackage[sort&compress]{natbib}
\usepackage{xcolor}

\newcommand{\beq}{\begin{equation}}
\newcommand{\eeq}{\end{equation}}
\newcommand{\bea}{\begin{eqnarray}}
\newcommand{\eea}{\end{eqnarray}}

\newcommand{\rvec}{{\bf r}}
\newcommand{\Rvec}{{\bf R}}

\begin{document}

\title{Static self energy and effective mass of the homogeneous electron gas\\
from Quantum Monte Carlo calculations}

\author{Markus Holzmann}
\author{Francesco Calcavecchia}
\affiliation{Univ. Grenoble Alpes, CNRS, LPMMC, 38000 Grenoble, France}
\author{David M. Ceperley}
\affiliation{University of Illinois Urbana-Champaign,Urbana, Illinois 61801, USA }
\author{Valerio Olevano}
\affiliation{CNRS, Institut N\'eel, 38042 Grenoble, France}

\date{\today}

\begin{abstract}
We discuss the methodology of quantum Monte Carlo calculations of the effective mass
based on the static self energy, $\Sigma(k,0)$. We then
use variational Monte Carlo calculations of $\Sigma(k,0)$
of the homogeneous electron gas at various densities to obtain results very close
to perturbative $G_0 W_0$ calculations for  values of the density parameter
$1 \le r_s \le 10$. The obtained values for the effective mass are close to 
diagrammatic Monte Carlo results and disagree with previous quantum Monte Carlo calculations
based on a heuristic mapping of excitation energies to those of an ideal gas.
\end{abstract}


\maketitle

Landau's Fermi liquid theory \cite{Landau1} has provided 
a paradigmatic frame for the phenomenological
description of equilibrium and transport properties of degenerate fermions
in terms of a very few characteristic parameters.
Silin \cite{Silin} has provided the path to generalize for long-range forces, in order to
extend it to normal metals in condensed matter \cite{NozieresLutt1,NozieresLutt2}.
Although the formal structure of the underlying microscopic theory has been known for a long time
\cite{Landau2,Luttinger,Nozieres}, 
most explicit calculations of the Fermi liquid
parameters rely on approximative, perturbative schemes \cite{vignale,book}.
As diagrammatic perturbation theory is not expected to converge for typical electronic
densities, basic Fermi liquid parameters 
of the 3D homogeneous electron gas (jellium), 
such as the effective mass $m^*$ and the renormalization factor $Z$, are sensitive to the underlying approximation \cite{Giuliani}.

Recently, variational diagrammatic Monte Carlo calculations (VDiagMC)
\cite{VDiagMC,Haule} for 3D jellium have been performed to include and control higher order terms
of the perturbation series. Those calculations found an overall reasonable agreement for 
$Z$
with previous quantum Monte Carlo calculations (QMC) \cite{Momk3D}.
However,
VDiagMC results on $m^*$
have been strongly questioned by QMC calculations of Ref.~\cite{Azadi}
yielding substantially different values.

In this paper, we revisit the methodology of zero temperature QMC calculations of the
effective mass, in order to resolve the discrepancy between QMC and perturbative/VDiagMC results, 
 show how such
calculations can be done and provide new results for 3D Jellium.
In principle, the effective mass can also be calculated from
the temperature dependence of thermodynamic quantities
\cite{mstarT,FiniteT2D}, e.g. from finite-temperature path-integral results
\cite{3DEGT,WDM,3DEGTX}; finite temperature methods will not be discussed here. In contrast to systems with 
short range interaction \cite{Shiwei}, size corrections are expected to play
an important role for charged systems \cite{FSE} as explained in detail below.

{\em Landau energy functional.}
Landau \cite{Landau1} phenomenologically characterized the low energy excitation of a Fermi liquid
by assuming a one-to-one correspondence of states of the ideal Fermi gas and those of the interacting system, such that elementary excitations of the interacting systems are still described
in terms of ideal gas occupation numbers before adiabatically switching on the interaction.
Changes of the total energy, $\delta E$, can then be considered as a functional
of changes in the quasi-particle occupation number $\delta n_{p \sigma}$ of the
momentum $p$ and spin quantum number $\sigma$
\beq
\delta E = \sum_{p \sigma} (\varepsilon_p+ \mu)  \delta n_{p \sigma}
 + \frac{1}{2V} \sum_{p \sigma ,p' \sigma'} f(p \sigma, p' \sigma')  \delta n_{p \sigma}  \delta n_{p' \sigma'} 
\label{Efunc}
\eeq
Here, $\mu$ is the chemical potential, $\varepsilon_p=(p-p_F)p_F/m^*$ is the 
quasiparticle energy which determines the effective mass for momenta in the vicinity
of the Fermi momentum, $p_F$, and $f(p \sigma, p' \sigma')$ is the quasi-particle
interaction, independent of volume, $V$, to leading order. Here, and in the following, we 
assume a homogeneous system with isotropic Fermi surface.

The success of Landau's Fermi liquid theory
started with its application to strongly interacting quantum liquids \cite{LFT}. 
Postulating an entropy functional in terms of quasi-particle occupations,
non-trivial predictions could be made using only a few
parameters, notably the effective mass as a coefficient of 
the quasi-particle energy $\varepsilon_p$.

Fermi liquid behavior results from the assumption
of certain analytical properties of 
fundamental correlation functions \cite{Nozieres},
notably the existence of a Fermi surface \cite{Luttinger} defined by the sharp discontinuity $Z$ 
of the momentum distribution at zero temperature, and the effective mass $m^*$ 
obtained from the
dispersion of the quasi-particle peak of the spectral function.

Both quantities, $Z$ and $m^*$, can thus be operationally defined from the single particle
Greens function, conveniently expressed in Fourier space, 
\bea
G(k,z)&=& G^+(k,z)+G^-(k,z)  
\label{LehmanG} \\
G^{\pm}(k,z)&=&\sum_n \frac{|\langle E_n^{N\pm 1} | a_k^\dagger |E_0^N \rangle|^2}{z-(\pm(E_n^{N \pm 1}-E_0^N)-\mu)}
\label{G+} 
\eea
where $|E_n^{N} \rangle$ denotes the $n^{th}$ eigenstate with
energy $E_n^{N}$ of the $N$-particle system. The self energy $\Sigma$ defined as
\beq
G^{-1}(k,z)=z+\mu -k^2/2m -\Sigma(k,z)
\label{invG}
\eeq
 captures all effects of
interactions where $m$ is the bare mass. 
The Fermi surface is then determined from $G^{-1}(k_F,0)=0$,
or $\mu=k_F^2/2m+\Sigma(k_F,0)$. Its analytic 
structure close to the singularity determines Fermi liquid behavior.
Under quite general assumptions 
the self-energy of the infinite system allows an expansion \cite{Nozieres}
\beq
\Sigma(k,z) -\Sigma(k_F,0) =
(k-k_F) \partial_k \Sigma(k_F,0)
+ z  \partial_z \Sigma(k_F,0)
\eeq
up to corrections of order $k^2$ and $z^2 \log z$.
The singularity dominating the Green's function close to the Fermi surface is then
\beq
G(k,z) \sim \frac{Z}{z-(k-k_F) k_F/m^*}
\label{Gquasip}
\eeq
with
\bea
Z^{-1} &=&1- \partial_z \Sigma(k_F,0)
\label{Zinv}
\\
\frac{m}{m^*} &=& Z \left(1+ \frac{m}{k_F}  \partial_k \Sigma(k_F,0) \right)
\label{mstarinv}
\eea
giving rise to a well defined quasi-particle behavior 
of strength $Z$ 
and energy $(k-k_F) k_F/m^*$.
Since $G(k,0)$ changes sign at $k=k_F$, the singularity of the Green's function 
close to the Fermi surface is entirely contained in either $G^+(k,z)$ or $G^-(k,z)$.
The real-time spectral function is obtained by approaching the real axis using $z=i\omega + \eta_k$
where $\eta_k=+0$ ($-0$) for $k>k_F$ ($k<k_F$).

Landau's energy functional may then be identified with the quasi-particle energies 
of the single particle propagator \cite{Luttinger,NozieresLutt2}, providing
a microscopic expression for the quasi-particle occupation number 
\cite{NozieresLutt2,Nozieres}. As knowledge of the quasi-particle energy is explicitly required and
its definition involves
off-diagonal matrix elements in the energy eigenstate representation,
this definition is purely formal and has not been of much practical use.
However, it provides a strong indication
that Landau's quasi-particle occupation number may not be expressible
as a simple static observable whose value can be determined from a single energy eigenstate. 
Only in the limit $k  \to k_F$, does the quasi-particle energy approximate
an exact energy eigenstate up to corrections of the order of the inverse lifetime \cite{Nozieres}, provided the thermodynamic limit is taken first.

Both, $Z$ and $m^*$, can be obtained from static 
observables at zero temperature. The value of the renormalization constant $Z$ can be read
off from the jump in the momentum distribution \cite{Momk3D}, whereas $\Sigma(k,0)$
can be obtained from the static response to an external perturbation 
$\xi (a_k + a_k^\dagger)$ as we will show below. Together they can be used to
calculate $m^*$ very near to the Fermi surface.

Landau's Fermi liquid theory successfully describes
thermal equilibrium or hydrodynamic transport observables \cite{LFT}, i.e. bulk properties. The form of Landau's energy functional, Eq.~(\ref{Efunc}), assures that its energy changes
with respect to variation of the quasi-particle occupations are to first order additive, with
corrections from a  small, $\sim 1/V$, pairwise interaction. 
Although these energy variations can be mapped to variations of the unperturbed 
ideal gas propagator within adiabatic perturbation theory \cite{Nozieres},
they cannot, in general, be mapped to the exact excited energy eigenstates
of the interacting system. 

The microscopic theory maps them to the single particle
quasi-particle spectrum, characterized by the emerging pole in the exact interacting 
propagator, Eq.~(\ref{Gquasip}), when
approaching the real axis, $z=i\omega + \eta_k$ with $\eta_k \to \pm 0$. 
However, for any finite system, the exact Green's
function, Eq.~(\ref{LehmanG}) is a highly irregular function on the real
axis; a smooth function can only be expected a finite
distance from the real axis, $|\eta_k| \gtrapprox k_F 2 \pi/(mL)$.
Instead, the effective mass formula, Eq.~(\ref{mstarinv}), involves only
static quantities with $z=0$ and are well defined on the real axis, even
before the thermodynamic limit is performed. Their calculations
may still suffer from important finite-size effects \cite{FSE}, but numerical
extrapolations will eventually converge to the infinite system size values.

Considering the generalized Hamiltonian $\widetilde{H}=\sum_N (H_N-\mu N)$, an external perturbation
 $\xi (a_k + a_k^\dagger)$ couples the ground state of the $N$ particle
systems to excitations containing $N \pm 1$ particles. 
From time-independent perturbation theory, restricting to states $|E^+ \rangle$ within
the subspace of $N$ and $N + 1$ particles,
the perturbed ground state up to linear order in $\xi$ can be written as
\beq
|E_k^+(\xi)\rangle= |E_0^N\rangle - \xi \sum_n \frac{|E_n^{N+ 1}\rangle \langle E_n^{N+ 1}| 
a_k^\dagger
|E_0^N \rangle}{E_n^{N+ 1}-E_0^N - \mu}
\eeq
yielding the energy to second order in $\xi$:
\beq
E_k^+(\xi)=E_0^N +\mu - \xi^2 \sum_n \frac{|\langle E_n^{N+1}| a_k^\dagger
|E_0^N \rangle|^2}{E_n^{N+1}-E_0^N-\mu}.
\eeq
Similarly $E_k^-$ is the ground state of the perturbed Hamiltonian 
restricted to the $N$ and $N-1$ subspaces.
The Green's functions are determined by comparing with the Lehmann representation, Eq.~(\ref{LehmanG}), 
\beq
G^{\pm}(k,0)=\lim_{\xi \to 0} \left[\pm (E_k^{\pm}(\xi)-E_0^N) -\mu\right]/\xi^2.
\eeq

Upper bounds to the ground state energies $E_k^\pm$ can be obtained with 
a variational ansatz for $|E_k^\pm \rangle$ and minimizing the expectation value
of the perturbed Hamiltonian
with respected to variational parameters.
Although technically a little bit more involved, calculations
of the static Green's function is thereby reduced
to a static response function, analogous to calculations of the density response \cite{Saverio}
previously employed using ground state Monte Carlo methods.

\begin{figure}
\includegraphics[width=9cm]{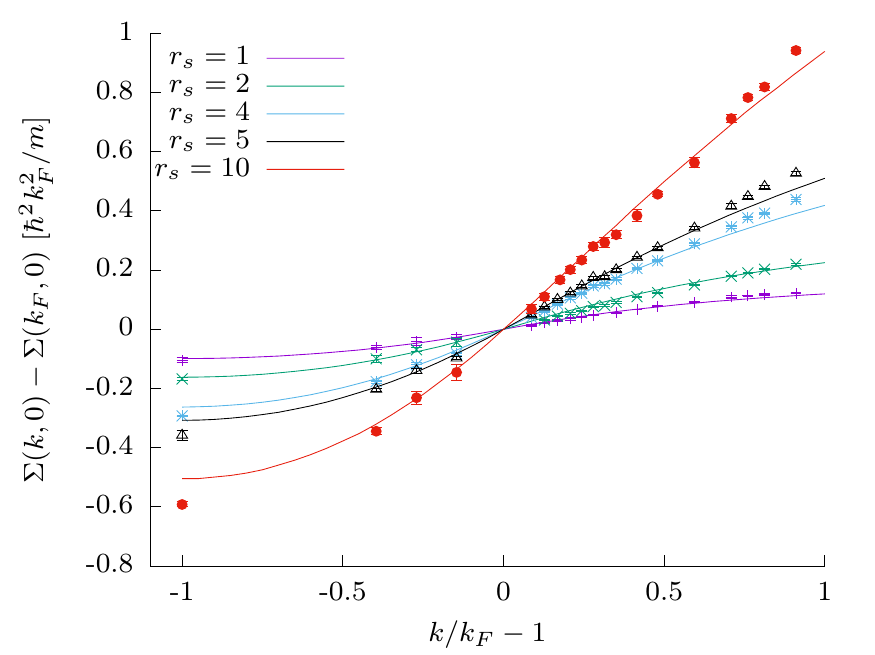}
\caption{Static self-energy for various densities ($r_s$) using backflow (BF) trial  wave functions and GC-TABC simulations  for $N=38$ electrons. They include size corrections. The color lines are from  $G_0 W_0$ calculations.
}
\label{SigmaBF}
\end{figure}

{\em Quantum Monte Carlo calculations}.
Let us now turn to the calculation of the static single-particle Green's 
function via quantum Monte Carlo methods, focusing on $G^+(k,0)$.
For that, we minimize the energy $E_T^+(\xi)$ of the generalized Hamiltonian
$\widetilde{H}$ using a trial wave function, $|\Psi_T(\xi) \rangle$, in the Fock space of $N$ and $N+1$ particle
wave functions providing an upper bound for $E_k^+(\xi)$.

Assuming that $\xi$ is sufficiently small, the  trial wave function can be expanded as
\beq
|\Psi_T(\xi) \rangle= |\Psi_0^N \rangle + \xi \sum_{i=1}^M \alpha_i |\Psi_i^{N+1} \rangle
\eeq
with $M$ the number of states in the basis.
It couples the ground state wave function $|\Psi_0^N \rangle$ of the $N$ particle system
(or our best variational ground state wave function)
with different wave functions $|\Psi_i^{N+1} \rangle $ of the $N+1$ particle states of total momentum
corresponding to $k$. The variational parameters are the set  $\{\alpha_i\}$.
A minimal choice consists in choosing $M=2$, with $|\Psi_1^{N+1} \rangle$ as a candidate for
a pure excited state wave function, minimizing separately the excited 
state energy $E_k^{N+1}$ in the $N+1$ section of momentum $k$, and
$|\Psi_2^{N+1} \rangle \sim a_k^\dagger | \Psi_0^{N} \rangle$; 
this should maximize
the overlap matrix elements of the perturbation with the ground state.  

Minimizing with respect to $\alpha_1,\alpha_2$  in the limit of $\xi \to 0$, we obtain a variational approximation for the
Green's function in the particle excitation sector
\bea
G^+_\mu(k,0) & =& -\frac{\zeta_1^2 \varepsilon_{22} - 2 \zeta_1 \zeta_2 \varepsilon_{12}
+ \zeta_2^2 \varepsilon_{11} }{\varepsilon_{11} \varepsilon_{22} - \varepsilon_{12}^2 }
\\
\text{with} \quad
\varepsilon_{ij} &=& \langle \Psi_i^{N+1} | H_{N+1} -E_0^N -\mu |\Psi_j^{N+1} \rangle
\\
\zeta_i&=&\langle \Psi_i^{N+1}|a_k^\dagger| \Psi_0^N \rangle 
\eea
where we have assumed normalized wave functions, e.g. $\langle \Psi_i^{N\pm 1} |\Psi_i^{\pm} \rangle =1$, with overall phases such that all matrix elements are real.

An analogous calculation in the hole sector  yields $G_{\mu}^-(k,0)$ from a variational
calculation based on  superposition of the lowest energy state for a hole excitation
and $a_k |\Psi_0^{N}\rangle$. Thus, the static Green's function $G_\mu(k,0)=G_{\mu}^+(k,0)+
G_{\mu}^-(k,0)$ is determined.

So far, the chemical potential, $\mu$,  entering as a parameter
in $G^\pm_\mu$, has not been specified yet.
Since single particle excitation are gapless in the Fermi liquid, 
the chemical potential can be fixed by the implicit equation
$\lim_{k \to k_F} G^{-1}_\mu (k,0)=0$. 

{\em Finite size effects.} Our Quantum Monte Carlo calculations are done for finite
number of electrons $N$ confined in a periodic cube of side $L$ and  volume $V=L^3$. Calculations must be extrapolated
to the thermodynamic limit. Shell effects in the single particle energy spectrum and
the Coulombic  interaction represent the main source of
finite size effects \cite{fse2}. 

\begin{figure}
\includegraphics[width=9cm]{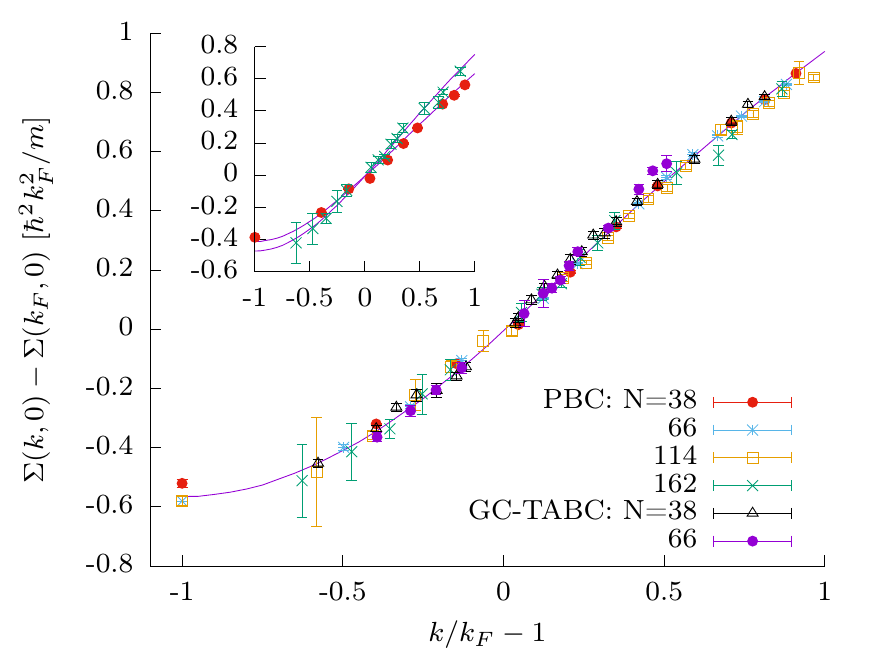}
\caption{Static self-energy for $r_s=10$ using SJ-VMC trial wave functions for simulations with
periodic boundary conditions (PBC) and GC-TABC for various sizes ranging from $N=38$
to $N=162$, size corrected according to Eq.~(\ref{deltaSigma}), the line is a fit to the
data.
The inset shows the uncorrected values for $N=38$ and $N=162$ (PBC), the lines
indicate the size corrections of the fit
based on Eq.~\ref{deltaSigma}).
}
\label{rs10Gamma}
\end{figure}

Shell effects 
 can be addressed by twisted 
boundary conditions \cite{TABC}
corresponding to a shifted grid calculation in momentum space.
Using grand-canonical twist averaging (GC-TABC) \cite{Momk2D} we obtain a sharp Fermi surface. We
spherically average $G_\mu^\pm(k,0)$ for any $k$ using $32$ 
equally weighted points that exactly integrates all polynomials on the
sphere up to the eighth order \cite{SphereWeb,Womersley}.

Although GC-TABC allows us to obtain $G_\mu^\pm(k,0)$ for arbitrary
$k$, size effects due to intrinsic two-body effects  remain. 
In charged systems, these are dominated by the long-range Coulomb interaction
\cite{fse1}. In particular, $z_k$ and $n_k$ are expected to suffer from important
size effects \cite{Momk3D} of order $1/L$. Instead of addressing them directly
which necessitates a thorough investigation as a function of $L$ and $k$, we will
determine the exact leading order form of the corrections from a diagrammatic analysis
based on Fermi liquid theory \cite{Nozieres}. Following Ref.~\cite{FSE}, within
the RPA approximation, $\delta \Sigma(k,0) = \Sigma_\infty(k,0)- \Sigma_N(k,0)$
is given by
\beq
\delta \Sigma(k,0)
\simeq
-\int_{-\pi/L}^{\pi/L} \frac{d^3 q}{(2 \pi)^3}
 \int_{-\infty}^\infty \frac{d \nu}{(2\pi)}
  \frac{v_{q}}{\epsilon(q,i \nu)} \frac{1}{i \nu +\mu-\varepsilon_{k+q}^0}
\label{dSigma}
 \eeq
where $v_q=4 \pi e^2/q^2$ is the Coulomb interaction, $\varepsilon_k^0=\hbar^2 k^2/2m$,
and the integral is restricted to a cube with
 $|q_\alpha|<\pi/L$ for any spatial component ($\alpha=x,y,z$).
Due to this restriction, we can use the expression $\epsilon(q,\omega) \simeq 
1-\omega_p^2/\omega$ for the dielectric function, where $\omega_p$ is the plasma frequency.
Since, the dominant contribution to the integral stems from the finite values of $\omega=i\nu$,
substituting this limiting form for the dielectric function captures the
exact behavior in the limit of small $q$.
The resulting integration then gives
\beq
\delta \Sigma(k,0) =C \frac{\varepsilon_k^0-\mu}{\omega_p +| \varepsilon_k^0-\mu |}
\label{deltaSigma}
\eeq
with $C \simeq 1.22 \, e^2/L$.
One can show that Eq.~(\ref{deltaSigma}) is indeed exact not only within RPA, if
$\varepsilon_k^0- \mu$ is replaced by the exact single particle energies
$(k-k_F) k_F/m^*$. This occurs since irreducible vertex corrections
approach $1/Z$ in the limit of vanishing momentum transfer at fixed frequency \cite{Nozieres}
and exactly cancel against the quasiparticle weight of the exact propagator replacing
the non-interacting propagator in the RPA expression.

\begin{table}
\begin{tabular}{ |c | l  l  l  l |}
\hline
$r_s$ & method & ~$Z$ & $m k_F^{-1}\partial_k \Sigma$ & $m^*/m$ \\
\hline 
$1$ &  BF-VMC & $0.86(1)$ \cite{Momk3D} & $0.17_{0.15}^{0.18}(1)$ & $1.00^{1.01}_{0.99}(1)$ \\
 &  SJ-VMC & $0.894(9)$ \cite{Momk3D} & $0.17^{0.18}_{0.15}(1)$ & $0.96^{0.97}_{0.95}(1)$ \\
 &  $G_0W_0$ (RPA) & $0.859$ \cite{Hedin} & $0.200$ & $0.970$ \cite{Giuliani}\\
 &  VDiagMC \cite{Haule} &  $0.8725(2)$ & $0.200(1)$ & $0.955(1)$ \\
\hline
$2$ & BF-VMC & $0.78(1)$ & $0.309^{0.361}_{0.280}(6)$  & $0.98_{0.94}^{1.00}(1)$  \\
    & SJ-VMC & $0.82(1)$ & $0.30^{0.31}_{0.28}(2)$ & $0.94^{0.95}_{0.93}(2)$ \\
 &  $G_0W_0$ (RPA) & $0.768$ \cite{Hedin} & $0.313$ & $0.992$ \cite{Giuliani}\\
 &  VDiagMC \cite{Haule} &  $0.7984(2)$ &   $0.328(4)$ &  $0.943(3)$ \\
\hline
$4$ & BF-VMC & $0.65(1)$ &  $0.538^{0.549}_{0.530}(7)$ &   $1.00^{1.01}_{0.99}(2)$       \\
    & SJ-VMC & $0.69(1)$ &  $0.55_{0.45}(2)$  &   $0.94^{1.00}(2)$        \\
 &  $G_0W_0$ (RPA) & $0.646$ \cite{Hedin} & $0.490$ & $1.039$ \cite{Giuliani}\\
 &  VDiagMC \cite{Haule} &  $0.6571(2)$ &  $0.528(5)$ & $0.996(3)$ \\
\hline
$5$ & BF-VMC & $0.59(1)$ &  $0.56^{0.65}(1)$ &   $1.09_{1.03}(3)$       \\
    & SJ-VMC & $0.61(1)$ &  $0.610^{0.624}_{0.596}(9)$  &   $1.02^{1.03}_{1.01}(2)$        \\
  &  $G_0W_0$ (RPA) & $0.602$ \cite{Hedin} & $0.569$ & $1.059$ \cite{Giuliani}\\
\hline
$10$ & BF-VMC & $0.41(1)$ &  $0.90^{0.98}_{0.88}(2)$ &   $1.28^{1.30}_{1.23}(3)$       \\
    & SJ-VMC & $0.45(1)$ &  $0.97^{1.03}_{0.91}(3)$  &   $1.13^{1.16}_{1.09}(3)$        \\
  &  $G_0W_0$ (RPA) & $0.45$ \cite{Hedin} & $0.98$ & $1.13$ \\
 \hline
\end{tabular}
\caption{Our QMC results with backflow (BF) and Slater-Jastrow (SJ) trial functions as compared to those of $G_0W_0$ (RPA) and
variational diagrammatic Monte Carlo (VDiagMC) \cite{Haule,KunPrivate,sigma}. 
Upper and low indices indicate systematic errors assuming different fitting functions/ranges
to determine $\partial_k \Sigma(k_F,0)$.
}
\label{tablefinal}
\end{table}

{\em Results and discussion.}
We have performed VMC calculations
for the 3D homogeneous electron gas based on analytical Slater-Jastrow (SJ) and
Slater-Jastrow backflow (BF) wave functions \cite{BF} as used in a previous study on the
renormalization factor \cite{Momk3D}. 
Its density, $n$, 
is parametrized by
$r_s \equiv a/a_B$, 
where  $a_B$ is the Bohr radius and $a=(4 \pi n/3)^{-1/3}$ is the
mean electron distance. Details of the VMC procedure are given in the Supplementary Material.

In 
Figure~{\ref{rs10Gamma} 
we illustrate the importance of size effects at $r_s=10$,
comparing canonical simulations with periodic boundary conditions (PBC)
from system sizes, $N=38$ to $114$ using SJ wave functions.
Although, the bare curves seem to indicate only
small variations with size, the size corrected curves
based on the analytical formula above show that the bare curves for such small systems
are still very far from reaching the thermodynamic limit. 
Due to the slow decay $\sim L$ of the corrections, 
we have not attempted any numerical extrapolation of the curves,. Extrapolation is more difficult
for smaller values of $r_s$ since variations are 
masked by the larger stochastic error.

In Figure~(\ref{SigmaBF}) we compare our size corrected results from BF-VMC calculations
using grand-canonical twist averaging to perturbative $G_0W_0$ results.
Even though $r_s=10$ is thought to be far outside the range of validity of a perturbative approach
our QMC results indicate only small modifications in the whole range $r_s \le 10$;
differences are hardly visible on the figures. 

In contrast to $Z$, perturbative calculations of $\partial_k \Sigma(k_F,0)$ 
seem to be much less sensitive to the underlying approximation scheme,
e.g. self-consistency and
vertex corrections \cite{GWKris,GWKotliar}.  We do not believe that the quantitative agreement 
of the static self energy between QMC and $G_0 W_0$
is a result of fortuitous error cancellations.

In order to deduce the effective mass,
we have fitted our QMC results for $\Sigma(k,0)$ around $k_F$ to obtain
$\partial_k \Sigma(k_F,0)$. In
table \ref{tablefinal} we summarize our results based on size corrected
GC-TABC calculations for 
$N=66$ SJ and $N=54$ BF wave functions. We see that
the decrease of $Z$ competes with the increase
of $\partial_k \Sigma(k_F,0)$, resulting in values of $m^*/m$ very close to one.
However, since $mk_F^{-1}\partial_k \Sigma(k_F,0)$ remains 
smaller than one even at $r_s=10$, the lowering of $Z$ with increasing $r_s$ eventually
dominates the effective mass and $m^*$ clearly increases for $r_s  \gtrsim 4$.

Similarly to $Z$, the change from the SJ trial function to more accurate BF trial functions 
reduces  $\partial_k \Sigma(k_F,0)$ by a small amount, slightly larger than our statistical
resolution. This provides a rough estimate of
the bias due to the trial wave function.
Since our approach is variational, we expect that the our results provide
upper bounds to $\partial_k \Sigma$. 
In addition, more correlated wave functions
tend to lower the values of $Z$ \cite{Momk3D,Max}, 
so that our  results for $m^*$ are likely lower bounds.
Future studies based on iterative backflow 
and machine learning wave functions \cite{BFegas,Max,MatthewEG} can be used to further reduce the wave function bias. 

Our results are in rather good agreement with perturbative $G_0W_0$ calculations
\cite{Hedin,Giuliani} 
and more recent variational digrammatic Monte Carlo calculations including higher order
diagrams \cite{Haule,KunPrivate}. They are at variance with
previous QMC calculations \cite{Azadi} of $m^*$ which are based on a heuristic mapping
of excitation energies to the ideal gas and not on the properties of the single particle Green's function. As we have reviewed above, 
the use of Landau's energy functional to determine Fermi liquid parameters from 
the excitation spectrum of finite systems is highly problematic. The comparison with those results is further detailed in the Supplementary Material.

The quantitative agreement between
two methodologically and numerically different methods, real space QMC and VDiagMC,
is highly encouraging. Comparisons with high precision measurements, as already
done 
in solid sodium \cite{Na} and lithium \cite{Li1,Li2}
for the renormalization factor, $Z$, 
can now be extended to the band width and effective mass.

\begin{acknowledgments}
 The authors acknowledge support from
the Fondation Nanosciences de Grenoble. DMC  is supported by the U.S. Dept. of Energy,  CMS program  DE-SC0020177.
M.H. thanks Saverio Moroni for many valuable discussions.
Computations were done 
using the
\href{(https://urldefense.com/v3/__https://gricad.univ-grenoble-alpes.fr}
{GRICAD}
infrastructure 
which
is supported by Grenoble research communities, and HPC resources 
from GENCI-IDRIS A0140914158. 
\end{acknowledgments}

\section*{Supplementary Material}
\renewcommand{\thetable}{S.\arabic{table}}
\renewcommand{\thefigure}{S.\arabic{figure}}
\renewcommand{\theequation}{S.\arabic{equation}}

In this Supplementary Material we provide details of the VMC calculations, how the self-energy is extracted, and how errors are estimated.
Then we present data for different sizes of the simulation cell  in table \ref{tablebrut}. Finally we compute an effective band mass of excitation energies and compare with a recent calculation.

\subsection{Variational Monte Carlo Simulations}

We have performed variational quantum Monte Carlo
simulations using both $\Gamma$ point  and grand-canonical twist-averaged boundary conditions
 (GC-TABC) conditions for systems containing between $N=38$ and $N=162$ electrons.
In these calculations we used 
Slater-Jastrow (SJ) and Slater-Jastrow-Backflow (BF)  trial wave functions
with analytical expressions for the Jastrow and backflow potentials as detailed
in Ref.~\cite{BF}  for the 3D jellium, and used in a previous
calculation for momentum distribution
\cite{Momk3D} and for
the renormalization factor $Z$ entering the effective mass formula.

\subsubsection{Wave function structure}

A SJ/BF wave function for the ground state is
\beq
\Psi_0^{N}
=  D^{N}_k e^{-U_{N}}
\eeq
with
\bea
D^{N} &=& \det_{qn} \varphi_q(r_n) \\
\varphi_q(r) &=& e^{i q \cdot r}
\eea
where $q$ contains $N$ wave vectors of lowest $|q|$ (here and in the following we do not
explicitly consider the spin-structure of our paramagnetic system where the determinant can
be further reduced to the product of two determinants, one for each spin), and
 $U_{N}$ denotes the symmetric pair correlation factor 
\beq
U_{N}= \sum_{i<j} u(r_i-r_j)
\eeq
In the case of SJ wave functions, the arguments of the orbitals $\varphi_q(\cdot)$ are the bare electron coordinates, whereas in backflow wave functions they are shifted by a many-body correlation factor \cite{BF}.

Our perturbed wave functions, Eq.~(12), is then built from the SJ (BF) ground state
wave function of the $N$ particle ground state, $\Psi_0^{N}$, and from $M=2$ non-orthogonal 
states, $\Psi_1^{N \pm 1}$ and $\Psi_2^{N \pm 1}$. 
To describe 
a perturbation of momentum $k$,  
$\Psi_{1/2}^{N \pm 1}$ describe wave functions where essentially
a plane wave orbital of wave vector $k$ is added/removed from the $N$ particle ground state.
The amount of correlation
changes between $\Psi_1^{N \pm 1}$ and $\Psi_2^{N \pm 1}$.
In the following, we describe both wave functions in detail.

Let us first discuss $\Psi_1^{N \pm 1}$ which is chosen in order to represent a good trial
wave function for an $N \pm 1$ particle energy state with momentum $ \pm k$ compared to the
ground state.
For $\Psi_1^{N \pm 1}$, we used the simplest extension of the ground state state wave function
to a correlated excited state 
which approximately describes an exact eigenstate of momentum $\pm k$ in the $N \pm 1$ particle sector
\beq
\Psi_1^{N \pm 1}
=  D^{N \pm 1}_k e^{-U_{N\pm 1}}
\eeq
In the case of particle excitations ($\Psi_1^{N+1}$), the orbitals of the Slater determinant
$D^{N +1}$
now contain 
also $k$ in addition to all wave vectors occupied in the ground state,
whereas for hole excitations ($\Psi_1^{N-1}$), the orbital of wave vector $k$ is removed.
Further, $U_{N \pm 1}$ denotes the symmetric pair correlation potential 
\beq
U_{N \pm 1}= \sum_{i<j} u(r_i-r_j)
\eeq
which is built from the same Jastrow potential $u(\cdot)$ used in the ground state,
and the summation is over all $N (N \pm 1)/2$ pairs of particles.

In contrast to $\Psi_1^{N \pm 1}$, where we have used a straightforward candidate for minimizing
the total energy, 
\beq
E_{11}^{N \pm 1}  \equiv \frac{\langle \Psi_1^{N \pm 1} | H_{N \pm 1} | \Psi_1^{N \pm 1} \rangle}{\langle \Psi_1^{N \pm 1} |\Psi_1^{N \pm 1} \rangle}
\eeq
$\Psi_2^{N\pm 1}$ is chosen to maximize the overlap with the perturbation acting on the
ground state,
$a_k^\dagger |\Psi_0^N\rangle $. The natural candidate is then to directly use
a state
$|\Psi_2^{N + 1} \rangle \propto a_k^\dagger |\Psi_0^N\rangle $ in the case of particle excitations,
and $|\Psi_2^{N - 1} \rangle \propto a_k |\Psi_0^N\rangle $ for holes.

In the coordinate representation, the unnormalized
hole state $\Psi_2^{N-1}$ can be written as
\beq
\Psi_2^{N-1}(\Rvec_{N-1}) = \int d \rvec_n e^{-i k \cdot r_N} \Psi_{0}^N (\Rvec_N)
\eeq
with $\Rvec_N=(\rvec_1,\dots,\rvec_N)$. 

In practice, we perform a VMC calculation of weight $|\Psi_1^{N-1}(\Rvec_{N-1})|^2$,
and matrix elements/ expectation values containing
$\Psi_0^{N}(\Rvec_N)$ and $\Psi_2^{N-1}(\Rvec_{N-1})$ can be accessed via 
introducing additional integrals over $\rvec_N$ and reweighting.

In the particle excitation sector, $\Psi_2^{N+1}$, for SJ wave functions, we can 
explicitly write down $|\Psi_2^{N+1}\rangle  \propto a_k^\dagger |\Psi_0^N \rangle$ in
the coordinate representation
\beq
\Psi_2^{N+1}(\Rvec_{N+1}) = 
\sum_n \frac{\delta D^{N+1}_k}{\delta \varphi_k(\rvec_n) } \varphi_k(\rvec_n) e^{-U_{N}^n(\Rvec_{N+1})}
\label{psi2a}
\eeq
where 
\bea
U_N^n(\Rvec_{N+1}) & \equiv&  \sum_{i<j,i,j\ne n} u(r_i-r_j) = U_{N+1}(\Rvec_{N+1}) - \widetilde{u}(r_n)
\\
\widetilde{u}(r) &=& \sum_{i=1}^{N+1} u(r_i-r) - u(0)
\eea
and we can write
\beq
\Psi_2^{N+1}(\Rvec_{N+1}) =\sum_n \frac{\delta D^{N+1}_k}{\delta \varphi_k(\rvec_n) } \varphi_k(\rvec_n) e^{\widetilde{u}(r_n)} e^{-U_{N+1}(\Rvec_{N+1})}
\eeq
Denoting
$\widetilde{\varphi}_k(r)=\varphi_k(r) \exp[\widetilde{u}(r)]$ for the orbital $k$, we can
write the wave function in form of a determinant times a symmetric correlation function
\beq
\Psi_2^{N+1}(\Rvec_{N+1}) =
\det_{q n}  \widetilde{\varphi}_q(r_n) e^{-U_{N+1}(\Rvec_{N+1})}
\label{psi2det}
\eeq
with $\widetilde{\varphi}_{q}(r)=\varphi_q(r)$ for $q \ne k$. Derivatives of the determinant
needed for the local energy
can be calculated similarly as backflow \cite{Kwon93,BFiter}.

In the case of BF, we have used Eq.~(\ref{psi2det}) where the bare coordinates are replaced
by the backflow coordinates constructed exactly as in the $N+1$ particle case. Although
this wave function is not exactly proportional to $a_k^\dagger |\Psi_0^N\rangle$, we
don't expect any significant differences, e.g.
 we have checked that
$ |\langle \Psi_2^{N+1} | a^\dagger_k |\Psi_0^N \rangle |^2$  coincides with $\langle \Psi_0 | a  a^\dagger_k |\Psi_0^N \rangle = 1-n_k$  within our stochastic error.

\subsection{GC-TABC}

In the GC-TABC procedure \cite{Momk2D,fse2,bandgap} the volume of the simulation box is fixed
by $V=N/n$ where $n$ is the electronic density. Each simulation is characterized by a twist vector; the phase that the
trial function picks up as an electron exits the supercell on one side and re-enters the other side. 
For a given twist vector, the Slater determinant
of our wave function consists of all plane wave states of wave vector $\bf k$ such that
$|{\bf k}| \le k_F$. The number of electrons then depends on the twist vector, $N_\theta$,
but  the number of electrons averaged over all twists equals $N$.
Expectation values of observables and matrix elements
are calculated independently for each twist angle and then averaged over twists.
In all our simulations $N$ has been chosen to be one that has a closed shell ground
state at the $\Gamma$ point.

Notice that all our trial wave functions have the properties that by space inversion they
equal their complex conjugate, hence the overlap integrals and matrix elements
are real.

For our study of the static self-energy, $\Sigma(k,0)$, for any given k and twist angle,
the observables
and matrix elements  $\varepsilon_{ij}$, $z_k$, and $n_k$, given by Eq.(14),(15), and (16)
as well as the ground state energy $E_0$
have been calculated from a variational Monte Carlo run with weight $|\Psi_1^{N+1} |^2$
using reweighting. (The equivalent procedure in the hole sector with $|\Psi_1^{N-1} |^2$.)
Within GC-TABC, the expectation values are
twist averaged before inserting them into the expression of the Greens function,
Eq.(13).
Note that Eq.(13) depends parametrically on the chemical potential $\mu$, but 
the Monte Carlo expectation values depend trivially on $\mu$.
For canonical VMC calculations at the $\Gamma$ point, values of $k$ are 
on a discrete grid given by the number of electrons $N$. Continuous values of $k$ are only
reached in the limit of $N \to \infty$.

For GC-TABC, any $k$ vector can be reached by imposing the corresponding twist angle. Spherical
averaging over all $k$ vectors of given magnitude then corresponds to averaging twist
over all twist angles yielding the same modulus of $k$.
In our GC-TABC calculations, we have used $32$ 
equally weighted points that exactly integrates all polynomials on the
sphere up to order eight \cite{SphereWeb,Womersley} for any given value of $k$.
Typically we have performed simulations for around $ \sim 10$ different k values
around $k_F$
in our runs for each $N$. 
In the GC-TABC runs we have chosen similar values of $k$ as
the canonical runs where the $k$ values are fixed by $N$. We also chose some intermediate ones
to interpolate better between them.
In our GC-TABC runs, we have avoided $k$ vectors too close to $k_F$, in particular
the region $|k-k_F| \lesssim \pi/L$  where no excitation would be possible in a canonical
simulation, since inside that region twist averaging can produce artifacts in the
self-energy. In any case, since the perturbed energies approaches $E_0\pm \mu$ 
for $k \to k_F$, the finite precision of our Monte Carlo evaluations additionally 
limit calculations approaching too closely the Fermi surface.

\subsection{ Estimation of Self-energy}

Given the value of the chemical potential, $\mu$, the Greens functions $G_\mu^{\pm}$
can be expressed in terms of QMC observables, e.g. Eq.(13-16) in the main text, (and thereby
 $\Sigma_\mu(k,0)$) by subtracting the single particle energies.
Its value at the Fermi surface,
$\Sigma_\mu(k_F,0)$,  as well as the slope of the self-energy have been obtained by
linear and quadratic fits
around $k_F$. The value of $\mu$ is then varied so that
$\Sigma_\mu(k_F,0)=\mu$ . In table \ref{tablebrut}, $\mu_N$ and $\partial_k \Sigma_N$
indicate the values obtained before size extrapolation. This procedure has been performed for the bare values at given system size, $N$,
and, equally, using the extrapolated self-energy $\Sigma_\mu^\infty(k,0)=\Sigma_\mu(k,0)+\delta \Sigma(k,0)$,
using Eq.~(18). In the table (which one), $\mu_\infty$ and $\partial_k \Sigma_\infty$ 
correspond to the chemical potential
and the slope  obtained using extrapolated self-energies extrapolated.

\begin{figure}
\includegraphics[width=9cm]{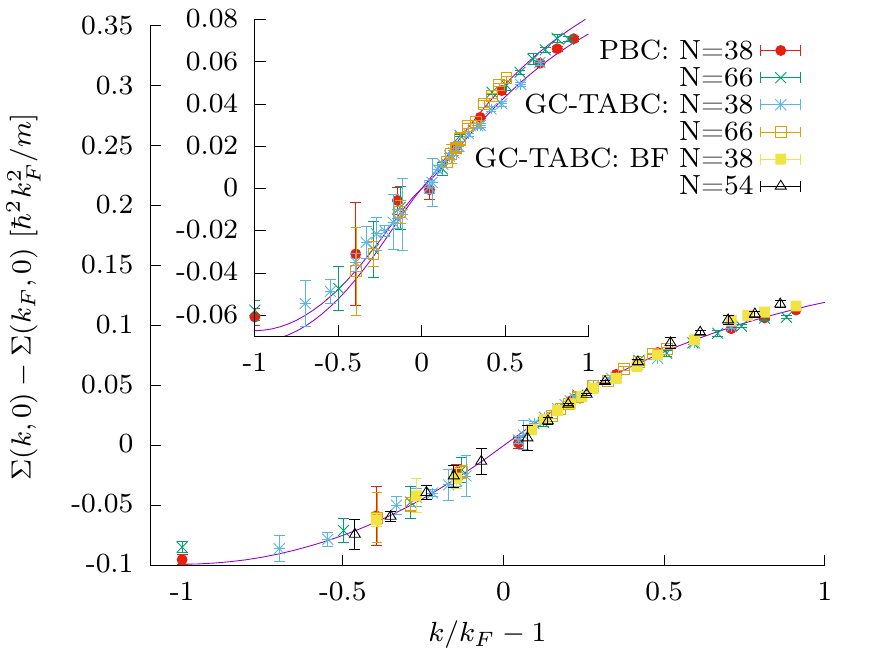}
\caption{Static self-energy for $r_s=1$ using SJ-VMC trial wave functions for simulations with
periodic boundary conditions (PBC) and GC-TABC for $N=38$ and $N=66$
together with those from BF-VMC trial wave functions 
	with GC-TABC $N=38$ and $N=54$. All curves in the main plot are
	size corrected according to Eq.~(17) of main text, the line is the $G_0 W_0$ self energy.
	The inset shows the uncorrected values (only SJ for clarity), the lines
	show the size corrections based on Eq.~(17) of the main text on the $G_0W_0$ curve.
}
\label{sigmars1}
\end{figure}
\subsection{ Estimation of Errors}

Since all matrix elements needed for the Greens function/ self-energy are calculated
within the same QMC simulations, the errors are not independent.
Therefore, we have estimated the final uncertainty of our self-energy, by determining
its spread over several independent calculations and different twist angles in GC-TABC.

The systematic errors for all quantities given in the tables
are estimated by looking at how the $\chi^2$ value varies with respect to the fitted value. 
Fits of GC-TABC curves usually yield values of $\chi^2$ close to one per degree of freedom. 
The $\chi^2$ value of the fits for canonical simulations at the $\Gamma$ point are
typically larger, in particular for
the $N=38$ and $N=66$ systems, because of the irregular filling of the shells of states,
and the fewer $k$ values close to $k_F$.

\subsection{ Finite Size Data}  

In the table of the main text, we have  
reported SJ and BF values from size extrapolated
GC-TABC using systems with $N=66$, and $N=54$, respectively.
Due to the systematic uncertainty of results with different sizes, we did not attempt
further numerical extrapolation.  Table  \ref{tablebrut} reports the data without size-corrections.

In Fig.~\ref{sigmars1} we further illustrate the changes due to size corrections for
the self-energy at $r_s=1$.

\begin{table*}
\begin{tabular}{|c|c|c| l l l l l |}
\hline
$r_s$ & $\Psi$ & BC & $N$ & $\mu_N$ & $ \mu_{\infty}$ & $mk_F^{-1}\partial_k\Sigma_N$  & $m k_F^{-1}\partial_k \Sigma_\infty$  \\
\hline 
\hline
1  & SJ & $\Gamma$     & 38 & $0.6224^{0.632}_{0.618}(8)$ & $0.628^{0.637}_{0.626}(3)$ 
& $0.095^{0.114}_{0.068}(3)$        & $0.18^{0.20}_{0.16}(1)$ \\ \cline{4-8}
   &    &              & 66 & $0.628^{0.632}_{0.623}(1)$ & $0.629^{0.641}_{0.627}(4)$
& $0.108^{0.112}_{0.091}(4)$        &   $0.18^{0.20}_{0.16}(2)$         \\ \cline{3-8}
   &    &  GC-TABC  & 38 & $0.632^{0.634}_{0.621}(1)$  & $0.634^{0.639}_{0.623}(1)$
& $0.089_{0.079}^{0.117}(4)$       &  $0.182^{0.22}_{0.14}(6)$   \\ \cline{4-8}
   &   &          &  66 & $0.634^{0.635}_{0.629}(3)$ & $0.635^{0.640}_{0.632}(2)$ 
&  $0.09^{0.11}_{0.08}(2)$ & $0.17^{0.18}_{0.16}(1)$  \\ \cline{2-8} 
  &  BF & GC-TABC & 38 & $0.628^{0.629}_{0.623}(1)$ & $0.633^{0.637}_{0.628}(2)$
& $0.086^{0.097}_{0.077}(8)$ & $0.17^{0.18}_{0.15}(1)$ \\ \cline{4-8}
  &   &  & 54 & $0.628^{0.629}_{0.619}(1)$ & $0.634^{0.637}_{0.622}(4)$
& $0.097^{0.112}_{0.078}(4)$ & $0.167^{0.179}_{0.157}(4)$ \\ 
\hline
2  & SJ & $\Gamma$     & 38 & $0.2082^{0.2090}_{0.2070}(5)$ & $0.2110_{0.2088}(5)$ 
& $0.190^{0.207}_{0.185}(4)$        & $0.334^{0.370}(4)$ \\ \cline{4-8}
   &    &              & 66 & $0.235^{0.241}_{0.222}(2)$ & $0.250^{0.251}_{0.230}(2)$
& $0.173^{0.218}_{0.150}(6)$        &   $0.260^{0.33}(6)$         \\ \cline{3-8}
   &    &  GC-TABC  & 38 & $0.230^{0.232}_{0.222}(1)$  & $0.234^{0.240}_{0.221}(1)$
& $0.168_{0.183}^{0.165}(5)$       &  $0.310^{0.336}_{0.300}(5)$   \\ \cline{4-8}
   &   &          &  66 & $0.233^{0.234}_{0.226}(2)$ & $0.237^{0.243}_{0.238}(4)$ 
&  $0.186^{0.190}_{0.167}(5)$ & $0.30^{0.31}_{0.28}(2)$  \\ \cline{2-8} 
  &  BF & GC-TABC & 38 & $0.226^{0.233}_{0.217}(2)$ & $0.233^{0.235}_{0.226}(2)$
& $0.15^{0.18}_{0.13}(1)$ & $0.28^{0.29}(1)$ \\ \cline{4-8}
  &   &  & 54 & $0.212^{0.215}_{0.206}(1)$ & $0.219_{0.209}(2)$
& $0.201^{0.227}_{0.161}(4)$ & $0.309^{0.361}_{0.280}(6)$ \\ 
\hline
4  & SJ & $\Gamma$     & 38 & $-0.660^{-0.656}(1)$ & $-0.655^{-0.654}_{-0.656}(1)$ 
& $0.312_{0.279}(3)$        & $0.511^{0.555}_{0.457}(3)$ \\ \cline{4-8}
   &    &              & 66 & $-0.638_{-0.650}^{-0.631}(7)$ & $-0.636^{-0.625}_{-0.646}(7)$
& $0.32^{0.37}_{0.28}(3)$        &   $0.50^{0.56}_{0.47}(3)$         \\ \cline{3-8}
   &    &  GC-TABC  & 38 & $-0.633^{-0.629}_{-0.644}(2)$  & $-0.628^{-0.622}_{-0.644}(1)$
& $0.32_{0.34}^{0.30}(1)$       &  $0.53^{0.57}_{0.51}(1)$   \\ \cline{4-8}
   &   &          &  66 & $-0.633^{-0.628}_{-0.646}(3)$ & $-0.632^{-0.623}_{-0.650}(4)$ 
&  $0.35^{0.37}_{0.31}(1)$ & $0.55_{0.45}(2)$  \\ \cline{2-8} 
  &  BF & GC-TABC & 38 & $-0.653^{-0.652}_{-0.661}(2)$ & $-0.649^{-0.645}_{-0.663}(2)$
& $0.285^{0.301}_{0.278}(8)$ & $0.51^{0.52}_{0.48}(1)$ \\ \cline{4-8}
  &   &  & 54 & $-0.668^{-0.661}_{-0.672}(3)$ & $-0.666^{-0.656}_{-0.671}(2)$
& $0.34^{0.36}_{0.31}(1)$ & $0.538^{0.549}_{0.491}(7)$ \\ 
\hline
5  & SJ & $\Gamma$     & 38 & $-1.1137^{-1.1110}_{-1.1172}(8)$ & $-1.1090^{-1.1067}_{-1.1134}(8)$ 
& $0.357^{0.370}_{0.309}(2)$        & $0.592^{0.637}_{0.530}(4)$ \\ \cline{4-8}
   &    &              & 66 & $-1.105^{-1.100}(2)$ & $-1.098^{-1.092}(2)$
& $0.40_{0.38}(1)$        &   $0.60_{0.57}(1)$         \\ \cline{4-8}
   &    &              & 114 & $-1.114^{-1.113}_{-1.117}(5)$ & $-1.113^{-1.108}_{-1.115}(5)$
& $0.43^{0.48}_{0.39}(3)$        &   $0.62^{0.66}_{0.58}(3)$         \\ \cline{3-8}
   &    &  GC-TABC  & 38 & $-1.089^{-1.085}_{-1.102}(3)$  & $-1.083^{-1.080}_{-1.102}(3)$
& $0.38^{0.41}_{0.36}(2)$       &  $0.62^{0.67}_{0.61}(2)$   \\ \cline{4-8}
   &   &          &  66 & $-1.085^{-1.082}_{-1.092}(3)$ & $-1.079^{-1.078}_{-1.097}(4)$ 
&  $0.415^{0.424}_{0.390}(9)$ & $0.610^{0.624}_{0.596}(9)$  \\ \cline{4-8} 
   &   &          &  114 & $-1.088^{-1.091}_{-1.072}(4)$ & $-1.080^{-1.047}_{-1.088}(4)$ 
&  $0.48_{0.43}(2)$ & $0.64^{0.65}_{0.56}(2)$  \\ \cline{2-8} 
  &  BF & GC-TABC & 38 & $-1.114^{-1.111}_{-1.127}(2)$ & $-1.105^{-1.104}_{-1.127}(2)$
& $0.35^{0.36}_{0.33}(1)$ & $0.573^{0.613}_{0.570}(5)$ \\ \cline{4-8}
  &   &  & 54 & $-1.118^{-1.114}_{-1.137}(3)$ & $-1.112_{-1.121}(2)$
& $0.35^{0.44}(1)$ & $0.56^{0.65}(1)$ \\ 
\hline
 10 & SJ  & $\Gamma$   & 38 & $-3.510_{-3.532}(1)$ & $-3.535^{-3.527}_{-3.539}(2)$ 
& $0.604_{0.544}(2)$  & $0.809^{0.891}(7)$  \\ \cline{4-8}
  &   &               & 66 & $-3.541^{-3.540}_{-3.555}(2)$ & $-3.535^{-3.527}_{-3.550}(2)$
& $0.61^{0.70}(1)$  & $0.91^{0.98}_{0.87}(1)$ \\ \cline{4-8}
  &   &               &114 & $-3.552^{-3.543}_{-3.572}(5)$ & $-3.540^{-3.535}_{-3.560}(5)$
&  $0.66^{0.69}_{0.56}(2)$  & $0.88^{0.92}_{0.81}(2)$ \\  \cline{4-8}
  &   &               &162 & $-3.563^{-3.533}_{-3.575}(10)$ & $-3.554^{-3.533}_{-3.571}(10)$
&  $0.70^{0.75}_{0.62}(5)$  & $0.91^{0.95}_{0.85}(5)$ \\  \cline{3-8}
  & &  GC-TABC  & 38 & $-3.514^{-3.507}_{-3.524}(5)$ &  $-3.509^{-3.504}_{-3.538}(5)$
& $0.71^{0.74}_{0.65}(3)$ & $1.08^{1.16}_{0.89}(3)$ \\ \cline{4-8}
  & &           & 66 & $-3.502^{-3.491}_{-3.535}(6)$  &  $-3.494^{-3.491}_{-3.522}(6)$
& $0.68^{0.73}_{0.65}(2)$ & $0.97^{1.03}_{0.91}(3)$  \\ \cline{2-8}
  & BF & GC-TABC & 38 & $-3.536^{-3.534}_{-3.555}(5)$  &  $-3.526^{0.518}_{-3.553}(7)$
& $0.58^{0.61}_{0.53}(2)$ & $0.93^{0.96}_{0.86}(2)$\\ \cline{4-8}
  &  &  & 54 & $-3.567^{-3.563}_{-3.580}(5)$  &  $-3.558^{-3.553}_{-3.576}(5)$
& $0.62^{0.68}_{0.57}(2)$ & $0.90^{0.98}_{0.88}(2)$\\
\hline
\end{tabular}
\caption{
Summary of various QMC calculations using Slater-Jastrow (SJ) and Slater-Jastrow backflow (BF)
wave functions ($\Psi $) for simulations using periodic ($\Gamma$) and
grand-canonical twist averaged (GC-TABC) boundary conditions (BC).
Here, $\mu_N$ 
is the chemical potential
in units of $\hbar^2k_F^2/2m$, 
$mk_F^{-1}\partial_k \Sigma_N$  the dimensionless slope of the self-energy at $k_F$,
both for calculations of finite volume $\sim N$; $\mu_\infty$ and $\partial \Sigma_\infty$ denote
the corresponding values obtained adding the leading finite size corrections due to the
Coulomb singularity.
The statistical uncertainty of the fit in the least significant digit
is indicated in parentheses, whereas upper and lower indices indicate systematic shifts due to 
fitting range and order of the polynomial fit.
Results were obtained from fitting with a linear function in a region $|k-k_F|/k_F \ge 0.2$, systematic errors by fitting with up to a quadratic function 
}
\label{tablebrut}
\end{table*}

\subsection{Comparison with Reference \cite{Azadi}}

In Ref.~\cite{Azadi} an apparent effective mass was determined assuming a mapping between
energy eigenstates and occupations of the orbitals used in SJ and BF wave functions in the spirit
of Landau's energy functional. As we have outlined in the main text, relating quasi-particle
energies to exact energy eigenstates of a finite system, though highly
intuitive, is problematic and without firm theoretical justification. Differences our our values
of $m^*$ 
compared to those given in Ref.~\cite{Azadi} are therefore mainly due to the different methodology
used here, and not to differences in the VMC wave functions, nor to the use of DMC
in Ref.~\cite{Azadi} as we will shown in the following.

\begin{figure}
\includegraphics[width=9cm]{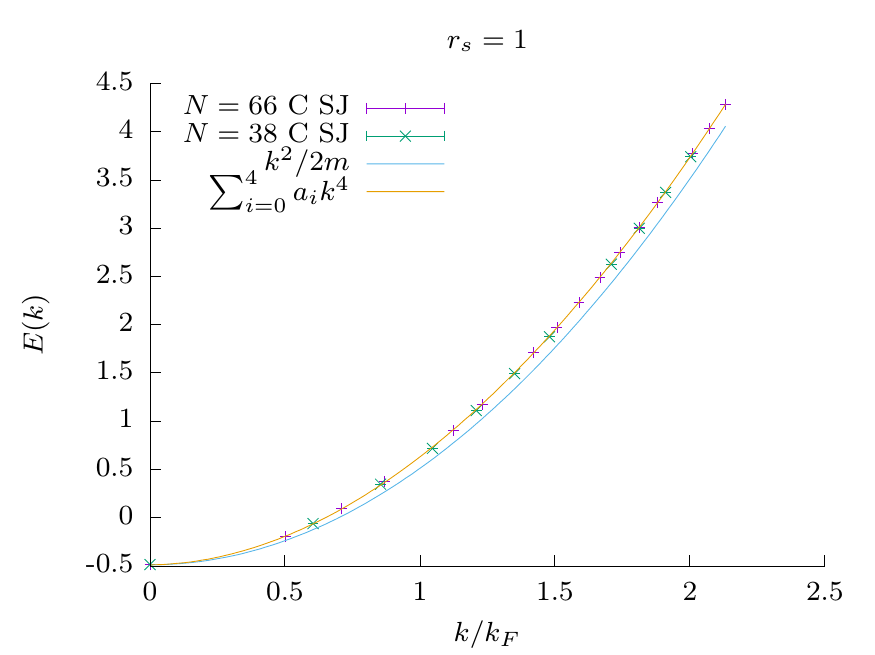}
\caption{Dispersion of the energy expectation values $E(k)$ based on $\Psi_1^{N \pm 1}$ for
SJ wave functions for $r_s=1$ using $N=38$  and $N=66$ electrons compared to the bare
$k^2/2m$ behavior of non-interacting electrons. The line through our data points is based
on a 4th order polynomial fit to the $N=66$ system. From the  fit at $k_F$ we obtain
$(k-k_F) k_F /m_{eff}$ with $m_{eff}/m \simeq 0.93$ close to the Fermi surface which is close to $0.915(1)$ given in Ref.~\cite{Azadi} extrapolated to infinite systems sizes.
}
\label{e_rs1}
\end{figure}
\begin{figure}
\includegraphics[width=9cm]{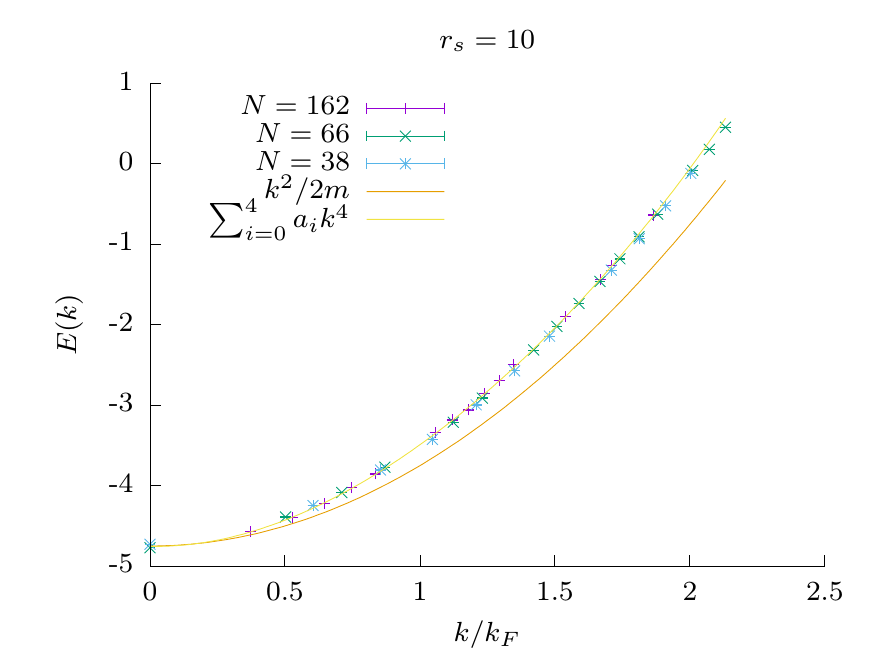}
\caption{Dispersion of the energy expectation values $E(k)$ based on $\Psi_1^{N \pm 1}$ for
SJ wave functions for $r_s=10$ using $N=38$, $N=66$, and $N=162$  electrons compared to the bare
$k^2/2m$ behavior of non-interacting electrons. The line through our data points is based
on a 4th order polynomial fit to the $N=162$ system. From the  fit at $k_F$ we obtain
$(k-k_F) k_F /m_{eff}$ with $m_{eff}/m \simeq 0.83$ close to the Fermi surface. The difference to $0.75(1)$ given in Ref.~\cite{Azadi} for the infinite system is compatible with the $1/N$ finite
size effects shown there.
}
\label{e_rs10}
\end{figure}

In our calculation of the static self energy, $\Psi_1^{N\pm 1}$ was chosen to provide a reasonable
approximation for an excited energy eigenstate of the $N \pm 1$ particle system with momentum $\pm k$. We can therefore directly compare the energy expectation value, and in particular the
''band'' dispersion
\bea
E(k) &=& \theta(|k|-k_F) \left[E_{11}^{N+1}(k) - E_0^N \right] 
\nonumber \\
&&+ \theta(k_F-|k|) 
\left[ E_0^N -E_{11}^{N-1}(k) \right]
\eea
with those given in Ref.~\cite{Azadi}. 

Fitting this energy dispersion
by a Pad{\'e} or polynomial fit,
the effective mass of Ref.~\cite{Azadi} was determined by the slope at the Fermi surface
\beq
E(k)= (|k|-k_F) k_F/m_{\text{eff}}, \quad k \to k_F
\eeq
where we denote $m_{\text{eff}}$ the thus obtained ''effective mass'' to avoid confusion with $m^*$ 
characterizing quasi-particle excitation energies.

In Figs.~\ref{e_rs1} and \ref{e_rs10}, we show the SJ band dispersion for $r_s=1$ and $r_s=10$.
In agreement with the results of Ref.~\cite{Azadi}, the dispersion of the band is steeper
than the one of non-interacting electrons. 
Using a fourth order polynomial fit for $E(k)$ of our largest systems, we get
$m_{\text{eff}}/m \simeq 0.93$ for $r_s=1$, $0.90$ for $r_s=2$, $0.88$ for $r_s=4$, $0.86$ for $r_s=5$,
and $0.83$ for $r_s=10$. Those values are consistent and in quantitative agreement 
with those of Ref.~\cite{Azadi} at comparable sizes. Optimization of excited state wave functions
as done in Ref.\cite{Azadi} as well as stochastic improvement via DMC make only a small change in $m_{\text{eff}}$, as already noticed in Ref.\cite{Azadi}.
Therefore the qualitative and quantitative difference of our values of $m^*$ in 
table I compared
to $m_{\text{eff}}$ 
calculations \cite{Azadi}
are due to the different methodology. The interpretation of $m_{\text{eff}}$ as effective mass
of quasi-particles invoking Landau's energy functional is problematic.


\begin{thebibliography}{99}
\bibitem{KunPrivate} Results given in \cite{Haule} are obtained at finite electronic temperature
$T=T_F/25$ where $T_F$ is the Fermi temperature 
and may be modified by zero temperature extrapolation.
\bibitem{sigma} VDiagMC and $G_0W_0$ values for $\partial_k \Sigma_\infty$ are
obtained using those given for $Z$ and $m^*/m$, assuming that the respective error is statistically independent.
\end{thebibliography}
\end{document}